\documentclass[reprint,longbibliography]{revtex4-1}

\pdfoutput=1

\usepackage[colorlinks,hyperindex,breaklinks]{hyperref}
\usepackage{graphicx}
\usepackage{amsmath}
\usepackage{bm}
\usepackage{bbold}
\usepackage{enumitem}
\usepackage{upgreek}
\setcitestyle{super,open={},close={}}

\begin{document}
\title{Dissipation and resonance frequency shift of a resonator magnetically coupled to a semiclassical spin}
\author{J. M. \surname{de Voogd}}
	\email{voogd@physics.leidenuniv.nl}
\author{J. J. T. \surname{Wagenaar}}
\author{T. H. \surname{Oosterkamp}}
\affiliation{Kamerlingh Onnes Laboratory, Leiden University, PO Box 9504, 2300 RA Leiden, The Netherlands}
\date{\today}

\begin{abstract}
We calculate the change of the properties of a resonator, when coupled to a semiclassical spin by means of the magnetic field. Starting with the Lagrangian of the complete system, we provide an analytical expression for the linear response function for the motion of the resonator, thereby considering the influence of the resonator on the spin and vice versa. This analysis shows that the resonance frequency and effective dissipation factor can change significantly due to the relaxation times of the spin. We first derive this for a system consisting of a spin and mechanical resonator and thereafter apply the same calculations to an electromagnetic resonator. Moreover, the applicability of the method is generalized to a resonator coupled to general two and more level systems, providing a key to understand some of the problems of two level systems in quantum devices.
\end{abstract}

\maketitle

Resonators and spins are ubiquitous in physics, especially in quantum technology, where they can be considered as the basic building blocks, as they can collect, store and process energy and information~\cite{nazarov_quantum_2009,you_superconducting_2005}. The validity of this information is, however, of limited duration as these building blocks leak practically always to the environment, which on its own can be seen as a bath of resonators and spins~\cite{caldeira_influence_1981,prokofev_theory_2000}. If in particular we focus on the situation where a resonator is coupled to a certain spin, then the spin's interaction with the environment naturally causes, besides a shift of resonance frequency, an extra dissipation channel for the resonator. Despite this simple qualitative explanation and many experimental~\cite{imboden_evidence_2009,venkatesan_dissipation_2010,bruno_reducing_2015} and theoretical efforts~\cite{caldeira_quantum_1983,sleator_nuclear_spin_1987,schlosshauer_decoherence_2008,pappas_two_2011}, an applicable full picture that quantitatively describes the response of a resonator coupled to a spin and their environments is still lacking. Here we derive classically the linear response function of the non-conservative system consisting of a resonator and a semiclassical spin. We show that the quality factor and resonance frequency of the resonator can be significantly influenced due to the relaxation times of the spin.\\
We start with a Lagrangian description, that includes the degrees of freedom of the resonator \emph{and} the spin, to find the coupled equations of motion (EOMs) that describe the resonator displacement and the spin magnetic moment, finding that this magnetic moment depends on the path the resonator takes. This is fundamentally different from conventional magnetic force microscopy (MFM)~\cite{rugar_magnetic_1990}, where one assumes a fixed polarization of the spins, like in magnetized samples. Even in magnetic resonance force microscopy (MRFM), which is usually focused on paramagnetic spins, it is generally assumed that the spin is not, or at least not significantly, influenced by the resonator~\cite{rugar_mechanical_1992,degen_nanoscale_2009,vinante_magnetic_2011}. We will show that this influence actually opens the dissipation channel and that the resonance frequency shift is more subtle than generally assumed.\\
Furthermore, we find in our analytical results that the interaction amplitude as function of temperature is a curve that for certain conditions shows an optimum, see Fig.~\ref{fig:figure3}, similar to the curves found in experiments where the tails have heuristically been fitted with power laws~\cite{imboden_evidence_2009,venkatesan_dissipation_2010}. Parts of the analysis we present here have been used by \citet{den_haan_spin-mediated_2015} to explain the experimental results obtained by approaching a native oxide layer on silicon with an ultra-sensitive MRFM probe. The equations derived in this paper were found to closely resemble the measured shift in resonance frequency and reduced quality factor as function of temperature and resonator - spin surface distance.\\
\begin{figure}
    \centering
    \includegraphics[width=\linewidth]{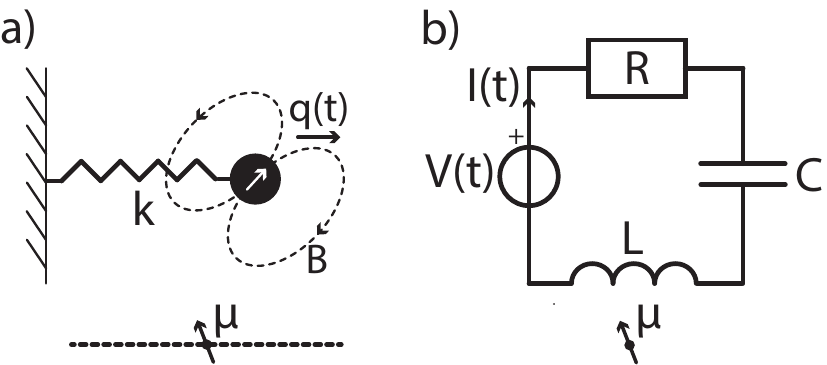}
    \caption{Schematic representation of spin $\bm{\mu}$ interacting with two types of resonators. a) The mechanical resonator with spring constant $k$ and displacement $q(t)$ of the magnet. The dashed line shows the position axis that is used in figure \ref{fig:figure2}. b) The electromagnetic resonator as a lumped element device. The current $I(t)$ through the inductor $L$ changes the magnetic field at the position of spin $\bm{\mu}$.}
    \label{fig:figure1}
\end{figure}
Although we start calculating the susceptibility of the more intuitive mechanical resonator, we will as well derive explicitly the (additional) impedance for electromagnetic resonators (see Fig.~\ref{fig:figure1}b versus \ref{fig:figure1}a, and Sec.~\ref{sec:EMres}), thereby making the results suitable for direct use by other fields in physics. Moreover, we will show the applicability of the theory to the case of the resonator coupled to two level systems (2LSs) and higher level quantum systems.\\
Finding an accurate description of the interaction of the building blocks of quantum devices with the environment can be seen as a widespread and major research area since not being able to understand, control and minimize the interaction is a major bottleneck in: the field of quantum computing~\cite{pappas_two_2011,bruno_reducing_2015}, detector fabrication in astronomy~\cite{day_broadband_2003,endo_chip_2013}, MRFM and high resolution MRI~\cite{poggio_force-detected_2010,kovacs_cryogenically_2005} and the development of optomechanical-like hybrid quantum devices~\cite{aspelmeyer_cavity_2014,lee_topical_2016}.\\

\section{Basic Principles}
\label{sec:bas}
The configuration of our theoretical analysis is given in Fig.~\ref{fig:figure1}a. A semiclassical spin, with magnetic moment $\bm{\mu}$, is located at laboratory position $\bm{r}_s$ and feels a magnetic field $\bm{B}(\bm{r}_s,t)$ that is produced by a magnet. The magnet is attached to a mechanical resonator that has spring constant $k$ and (effective) mass $m$. The origin of the laboratory frame is chosen to be the equilibrium position of the magnet's center. The displacement of the magnet from this equilibrium position is denoted by $q(t)$. See Fig.~\ref{fig:figure1}. The Lagrangian for this system is given by
\begin{equation}
L = \frac12m\dot{q}^2-\frac12kq^2+\bm{\mu}\cdot\bm{B}(q)+I_S.
\label{eq:LagS}
\end{equation}
$I_S$ stands for an expression with the internal spin degrees of freedom that needs to be included to derive the spin EOM. A more detailed account is left in App.~\ref{sec:appA}.\\
The resonator-spin system does not live in an isolated world. Therefore we include dissipation and decay to the environment into the EOMs. The first differential equation, derived with respect to the resonator displacement, includes the Raleigh dissipation $-\gamma\dot{q}$ of the resonator. This results in
\begin{equation}
\label{eq:harm}
m\ddot{q}+\gamma\dot{q}+kq-\bm{\mu} \cdot \frac{\partial}{\partial q}\bm{B}=F_{ext}(t),
\end{equation}
where the last term, $F_{ext}(t)$, is an external force that is exerted on the resonator.\\
Starting with the Lagrangian, which contains the degrees of freedom for the resonator \emph{and} the spin, leads to the force interaction term $-\bm{\mu}\cdot\frac{\partial}{\partial q}\bm{B}$. This is the same as $-\bm{\mu}\cdot\nabla B_{\parallel q}$, because of the vanishing curl of the magnetic field in free space. Here $\nabla\bm{B}_{\parallel q}$ is the gradient of the magnetic field component in the direction of the movement of the resonator. In MRFM $-\bm{\mu}\cdot\nabla B_{\parallel q}$ is often derived from calculating the force-field from the gradient of the potential energy $\nabla\left(\bm{\mu}\cdot\bm{B}\right)$, assuming that $\bm{\mu}$ does not depend on the position of the resonator~\cite{berman_magnetic_2006}. However, as $\bm{\mu}$ follows the classical path, we will show by solving the spin EOM that $\bm{\mu}$ is influenced by the resonator and it is therefore a priori not at all obvious that $\frac{\partial}{\partial q} \bm{\mu}=0$ as long as the spin degrees of freedom are not defined.\\\\

The other set of differential equations can be found by deriving the EOM with respect to the spin degrees of freedom. Since the spin interacts with the environment, we can expect an effectively decaying amplitude that is often described by $T_1$ and $T_2$; the time constants associated with the decay of the semiclassical magnetic moment longitudinal and perpendicular to the magnetic field, respectively~\cite{bloch_nuclear_1946}. If one assumes that the system consists of an ensemble of paramagnetic spins, instead of one,the average magnetic moment per spin decays to a certain equilibrium vector $\bm{\mu}_\infty$, according to the master equation~\cite{slichter_principles_1990}. However, if a single spin over time has on average the same behavior as the average of an ensemble at a certain moment, i.e. the spin satisfies ergodicity, then we can combine the ensemble's master equation and the single spin EOM to find a differential equation that describes the average behavior of the single semiclassical spin. This is the Bloch equation:
\begin{equation}
\dot{\bm{\mu}}=\gamma_s\bm{\mu}\times\bm{B}+T^{-1}\left(\bm{\mu}_\infty-\bm{\mu}\right).
\label{eq:Bloch}
\end{equation}
Here $\gamma_s$ is the gyromagnetic ratio and $T^{-1}\equiv\frac{1}{T_2}\left(\mathbb{1}-\bm{\hat{B}}\bm{\hat{B}}^T\right)+\frac{1}{T_1}\bm{\hat{B}}\bm{\hat{B}}^T$, where the hat denotes the unit-vector in the direction of the specified vector.\\
The spin equilibrium magnetic moment $\bm{\mu}_\infty(t)$ is the vector to which the spin magnetic moment would decay to if given the time. As the resonator moves, the magnetic field changes, and so does $\bm{\mu}_\infty$. We will assume that the environment of the spin is a heat bath, connected to the spin by means of the relaxation times. However, does the spin's equivalent spin ensemble have a well defined temperature? As derived in the original paper of \citet{bloembergen_relaxation_1948}, the differential equation describing the population difference $n$ for particles in a two level system is
\begin{align}
\frac{\mathrm{d}n}{\mathrm{d}t} = -2Wn+\frac{n_0-n}{T_1},
\end{align}
where $W$ is the probability rate that the particle changes energy level due to an applied field and $n_0$ is the population difference between the energy levels when the ensemble has the temperature of the heat bath. In other words $-2Wn$ is proportional to the incoming energy and $\frac{n_0-n}{T_1}$ is the connection to the heat bath. This results in
\begin{align}
n_\infty = \frac{n_0}{1+2WT_1},
\end{align}
where $n_\infty$ is the steady state solution. Thus when $2WT_1\ll1$ the spin ensemble, and hence our semiclassical spin, is connected well enough to the heat bath to assume that our spin has a well defined temperature. For spin-$\frac12$ this condition yields~\cite{bloembergen_relaxation_1948}
\begin{align}
\label{eq:condition}
\pi\gamma_s^2\left|\bm{B}'\right|^2q^2T_1g\left(\omega\right)\ll1,
\end{align}
where $\bm{B}'=\left.\frac{\partial}{\partial q}\bm{B}\right|_{\bm{r}=\bm{r}_s}$ and $g\left(\omega\right)$ the spin's normalized absorption line that is usually described by a Lorentzian or Gaussian that peaks around the Larmor frequency. This makes this condition hard to satisfy when the resonator has a resonance frequency around the Larmor frequency, and one should minimize the resonator's movement $q$. When this condition is not met, the spin saturates and the temperature increases or might be undefined~\cite{slichter_principles_1990}. However, for example in MRFM, mechanical resonators tend to have resonance frequencies much lower than the Larmor frequency and so it is much easier to satisfy this condition.\\
Assuming the condition is satisfied we can now derive $\bm{\mu}_\infty$ from the canonical ensemble and find for spin-$\frac12$
\begin{equation}
\bm{\mu}_\infty=\mu_s\tanh\left(\beta \mu_s\left|\bm{B}\right|\right)\bm{\hat{B}},
\end{equation}
where $\beta\equiv\frac{1}{k_BT}$ is the inverse temperature and $\mu_s\equiv S\hbar\gamma_s$ is the magnitude of the non-averaged spin magnetic moment with spin number $S=\frac12$. This result can easily be generalized for other spin numbers as is done in  App.~\ref{sec:appB}. For simplicity we will stick to the formula for spin-$\frac12$ particles here.\\

\section{Susceptibility}
\begin{figure}
    \centering
    \includegraphics[width=\linewidth]{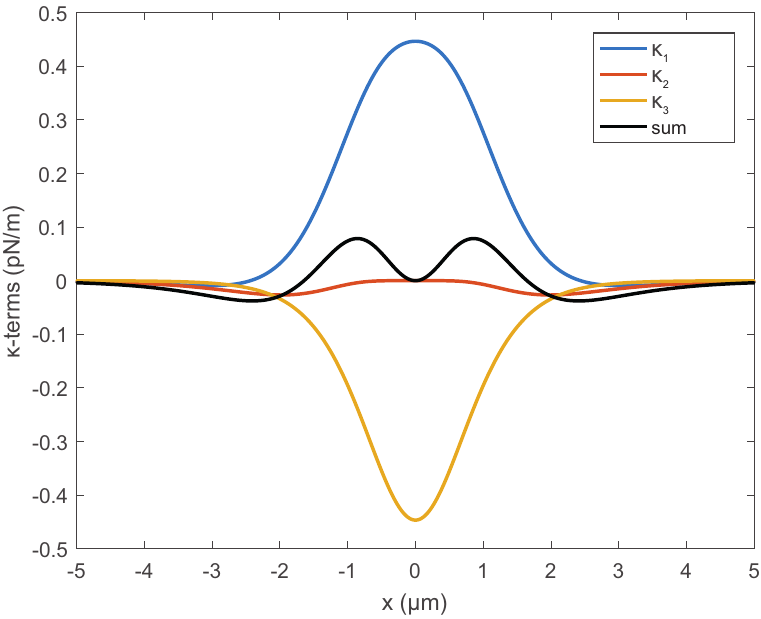}
    \caption{This graph shows the single spin contribution to the spring constant as function of a position axis parallel to the direction of resonator movement, as visualized by the dashed line in figure \ref{fig:figure1}a). In the simulation we attached a magnetic dipole (with magnetic moment of $19\ \text{pAm}^2$ in the direction of $q$) on a mechanical resonator. The resonator is connected, by means of the magnetic field, to an electron spin at a temperature of $300$ mK. The distance between the center of the dipole and $x=0$ is $2.5\ \upmu$m. To demonstrate the spatial behavior of the $\kappa$-terms we avoided imaginary terms by setting $T_1=0$ in $\kappa_2$ and $T_2=0$ in $\kappa_3$. The solid line shows the sum of these $\kappa$-terms.}
    \label{fig:figure2}
\end{figure}
To find the resonance frequency and quality factor of the resonator, we will need to calculate the interaction term up to linear order in $q$. Higher order terms will give rise to nonlinear effects. Interaction terms with even powers in $q$ are usually experimentally uninteresting since they will produce even multiples of the fundamental resonance frequency. These multiples are not measured or can easily be filtered. Uneven powers of $q$ can, however, lead to disturbing nonlinear effects like Duffing~\cite{kaajakari_nonlinear_2004}. One can lower the amplitude of $q$ to suppress higher order terms and therefore the nonlinear effects, but in experiments this is usually limited by the signal-to-noise ratio.\\
The zeroth order term does not contribute to the dynamics of the system, however it does give rise to a constant deflection of the resonator. This can be solved by shifting the origin of the laboratory frame by the amount of the deflection; this causes, however, a (usually small) change of the coordinates of the spin. We will provide an estimate of the deflection in App.~\ref{sec:appC} and leave it further out of account.\\\\
To find the interaction term $-\bm{\mu}\cdot\frac{\partial\bm{B}}{\partial q}$ up to first order in $q$, we need to solve Eq.~\ref{eq:Bloch} and find the constant and $q$-dependent parts. By substituting $q\to\lambda q$ we use perturbation theory to find
\begin{align}
\label{eq:inttermexpansion}
-\bm{\mu}\cdot\frac{\partial\bm{B}}{\partial q} = \bm{\mu}_0\cdot\bm{B}' + \lambda\left(\bm{\mu}_1\cdot\bm{B}'-q\bm{\mu}_0\cdot\bm{B}''\right)+\mathcal{O}\left(\lambda^2\right),
\end{align}
where $\bm{B}'=\left.\frac{\partial}{\partial q}\bm{B}\right|_{\bm{r}=\bm{r}_s}$ was defined previously and $\bm{B}''=\left.\frac{\partial^2}{\partial q^2}\bm{B}\right|_{\bm{r}=\bm{r}_s}$. Here $\bm{\mu}$ is perturbed into a $q$-independent part $\bm{\mu}_0$ and a linear term $\bm{\mu}_1$. The higher order terms $\mathcal{O}\left(\lambda^2\right)$ can be omitted, as well as the first term on the right hand side that only gives rise to the constant deflection.\\
At first we are mostly interested in solutions that do not decay over time and do not depend on initial conditions because then the linear response function can conveniently be given in the Fourier domain which makes it easy to compare with experiments. The Fourier Transform $\mathcal{F}\!\!\left\{\ \right\}$ of the linear response function, or simply susceptibility $\chi\left(\omega\right)\equiv\frac{\tilde{q}(\omega)}{\mathcal{F}\left\{F_{ext}\right\}}$, can be calculated from Eq.~\ref{eq:harm}
\begin{align}
\chi\left(\omega\right)=\frac{1}{k-m\omega^2+i\gamma\omega+\kappa},
\end{align}
where $\kappa=\kappa_1+\kappa_2+\kappa_3$, with $\kappa_1\equiv-\bm{\mu}_0\cdot\bm{B}''$ and $\kappa_2+\kappa_3\equiv\frac{\mathcal{F}\left\{\bm{\mu}_1\cdot\bm{B}'\right\}}{\tilde{q}(\omega)}$. Appendices~\ref{sec:appC},\ref{sec:appD} present the calculation of the $\kappa$-terms, which turn out to be:
\begin{align}
\kappa_1&=-\mu_s\tanh\left(\beta\mu_sB_0\right)\left|\bm{B}''_{\parallel\bm{\hat{B}}_0}\right|,\\
\kappa_2\left(\omega\right)&=-\frac{\mu_s}{B_0}\frac{\beta\mu_sB_0}{\cosh^2\left(\beta\mu_sB_0\right)}\left|\bm{\hat{B}}'_{\parallel\bm{\hat{B}}_0}\right|^2\frac{1}{1+i\omega T_1},\\
\kappa_3\left(\omega\right)&=-\frac{\mu_s}{B_0}\tanh\left(\beta\mu_sB_0\right)\left|\bm{B}'_{\perp\bm{\hat{B}}_0}\right|^2\cdot\nonumber\\
&\quad\left(1-\frac{2\frac{T_2}{T_1}-\left(\omega T_2\right)^2+i\omega T_2\left(1+2\frac{T_2}{T_1}\right)}{\left(1+i\omega T_2\right)^2+\left(\omega_sT_2\right)^2}\right),
\end{align}
where $\bm{B}_0\equiv\bm{B}\left(q=0\right)$ and the notation $\bm{v}_{\parallel\bm{\hat{B}}_0}$ and $\bm{v}_{\perp\bm{\hat{B}}_0}$ is used to indicate the part of $\bm{v}$ parallel and perpendicular to $\bm{\hat{B}}_0$ respectively for any vector $\bm{v}$. $\kappa_2$ and $\kappa_3$ are derived from $\bm{\mu}_{1_{\parallel\bm{\hat{B}}_0}}$ and $\bm{\mu}_{1_{\perp\bm{\hat{B}}_0}}$ respectively.\\
If we compare this result with the conventional approach that neglects the effect of the resonator on the spin, we see that in that approach we have only the term $\kappa_1$~\cite{garner_force_gradient_2004}. However, $\kappa_1$ is real and therefore it cannot describe the extra dissipation channel that has been seen in experiments~\cite{vinante_magneticv2_2011}. The derivation which has been done here does include the linear effect of the resonator on the spin and vice versa. This produces two extra terms in the linear response function that are partly imaginary. Each of the $\kappa$-terms is shown separately in figure \ref{fig:figure2} as a function of the spin position. This position axis is indicated in figure \ref{fig:figure1} by the dashed line. Which effect these terms have in practice, where usually more than one spin is present, will be shown in the next section.

\section{Spin Bath - Resonator Coupling}
\begin{figure}
    \centering
    \includegraphics[width=\linewidth]{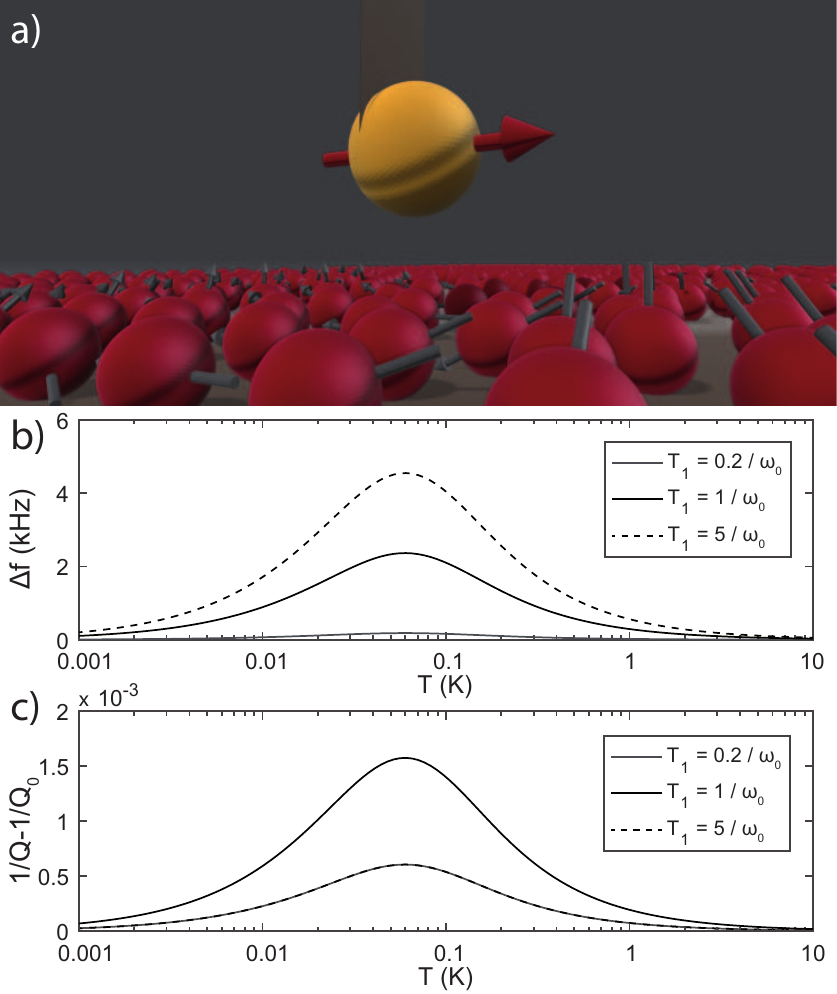}
    \caption{Calculated frequency shift and added dissipation of a mechanical resonator due to dangling bonds on a silicon surface, equivalent to the setup of~\citet{den_haan_spin-mediated_2015}. a) Impression of a NdFeB magnet (with magnetic moment $19\ \text{pAm}^2$ in the direction of $q$) attached to an ultrasoft silicon cantilever with spring constant $k=70\ \upmu$N/m, together leading to a natural frequency of $\frac{\omega_0}{2\pi}=3$ kHz. The center of the magnet is positioned at a distance of $2.2\ \upmu$m to the silicon sample. The surface of the sample has a native oxide containing $0.14$ electron spins/nm$^2$ that are visualized by the red balls (not to scale). The graphs b) and c) show the resonance frequency shift and the damping of the cantilever. The results are shown for various $T_1$, showing a maximal opening of the additional dissipation channel for $T_1=1/\omega_0$.}
\label{fig:figure3}
\end{figure}
\label{sec:bath}
We assume that all spins in the system act individually and do not influence each other, except through the relaxation times. We can then sum over the $\kappa$-terms for each spin to find the susceptibility of the resonator connected to a whole ensemble of spins, i.e. $\kappa=\sum_s\kappa_1(\bm{r}_s)+\kappa_2(\bm{r}_s)+\kappa_3(\bm{r}_s)$. Moreover, if the spins in the sample have an average nearest neighbor distance smaller than the typical spatial scale of the applied magnetic field, we can see the sample as a spin continuum and hence, instead of summing, integrate over the sample with spin density $\rho(\bm{r})$.\\
If we calculate the result for a volume with constant spin density, it can be found by partial integration of the volume in the direction of the movement of the resonator
\begin{align}
\label{eq:kbath}
\kappa(\omega) = &\rho\beta\mu_s^2C\frac{\left(\omega T_1\right)^2+i\omega T_1}{1+(\omega T_1)^2}\ +\nonumber\\
& \text{boundary term} + \mathcal{O}\left(\frac{1}{\left(\omega_s^2-\omega^2\right)T_2^2}\right),
\end{align}
with
\begin{equation}
C=\int_\mathcal{V}d^3\bm{r}\frac{\left|\bm{B}'_{||\bm{B}_0}\right|^2}{\cosh^2\left(\beta\mu_sB_0\right)}.
\end{equation}
The boundary term vanishes when the volume boundaries in the $\bm{q}$-direction are large. The $\mathcal{O}\left(\frac{1}{\left(\omega_s^2-\omega^2\right)T_2^2}\right)$ can be neglected for resonance frequencies away from the Larmor frequency and for $T_2\gg\frac{1}{\omega_s}$.\\
From $\kappa$ we can calculate the frequency and Q-factor shifts as seen in experiments by~\citet{den_haan_spin-mediated_2015}. For $Q_0\equiv\frac{\sqrt{km}}{\gamma}\gg\frac{1}{\sqrt{2}}$ the susceptibility has a maximum around the natural frequency $\omega_0\equiv\sqrt{\frac{k}{m}}$. Then, as long as the influence of the spin leads only to a small correction of the susceptibility, i.e. $\kappa\ll k$, the relative frequency shift is given by 
\begin{equation}
\frac{\Delta\omega}{\omega_0}\approx\frac12\frac{\text{Re}\left(\kappa(\omega_0)\right)}{k}.
\end{equation}
The imaginary part of $\kappa$ causes the change in Q-factor. The new Q-factor is given by
\begin{equation}
\frac1Q\approx\frac1Q_0+\frac{\text{Im}\left(\kappa(\omega_0)\right)}{k}.
\end{equation}
In Fig.~\ref{fig:figure3} we show an example of an experiment with a magnet attached to an ultrasoft cantilever, which is positioned above a silicon sample. The native oxide contains electron spins that interact with the resonating magnet. The frequency shift and quality factor depend differently on $T_1$. In this simulation we have set $T_2$ to zero only after we checked that the $\mathcal{O}$ term in Eq.~\ref{eq:kbath} can indeed be neglected: setting $T_2=T_1$ gives an additional frequency shift of about $1$ nHz and a five orders of magnitude lower shift in Q-factor compared to the results shown in Fig.~\ref{fig:figure3}c.

\section{Spin - Electromagnetic Resonator}
\begin{figure}
    \centering
    \includegraphics[width=\linewidth]{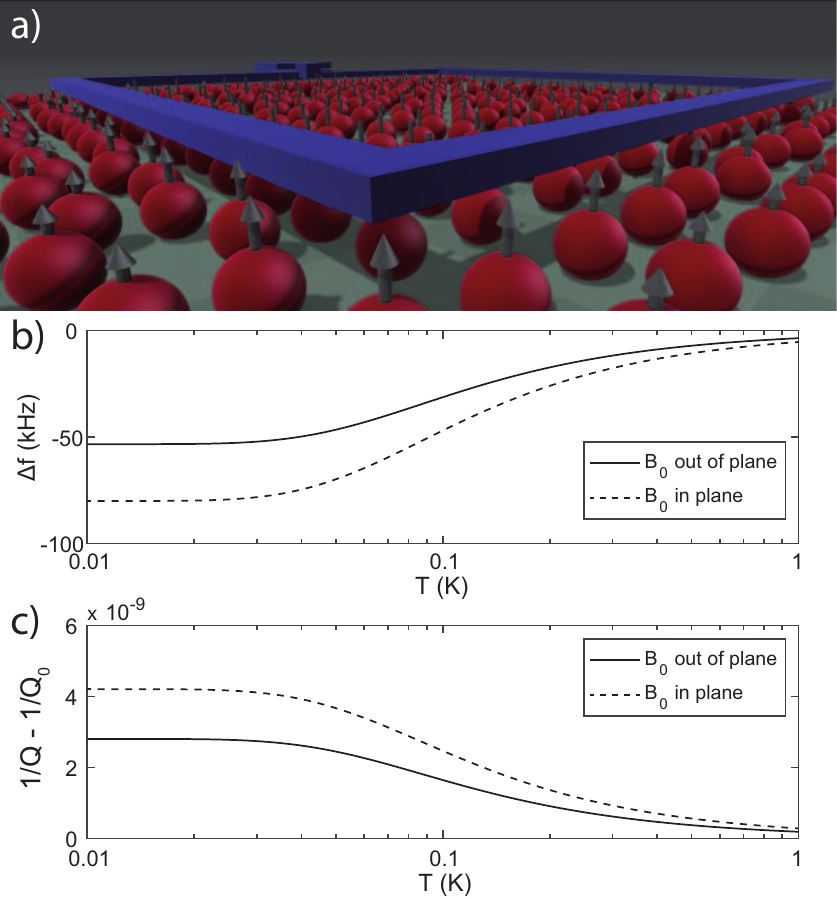}
    \caption{Simulation of frequency shift and added dissipation of an electromagnetic resonator due to dangling bonds at the sample's surface. a) Impression of an RLC-circuit with $10$ GHz natural frequency and $0.25$ nH inductance that consists of a $50\ \upmu$m $\times\ 50\ \upmu$m square which is positioned $50$ nm above a surface with $0.14$ electron spins/nm$^2$ b,c) Calculated results for a static external magnetic field of $0.1$ T that is oriented out of plane (solid curve) and in plane (dashed curve). For this simulation we assumed $T_2=0.01\ \upmu$s.}
    \label{fig:figure4}
\end{figure}
\label{sec:EMres}
In this section we calculate the complex impedance coming from a spin interacting with an electromagnetic resonator. The derivation is very similar to the mechanical resonator and hence we will largely copy the results. We will assume that the system can be described by a lumped element model, which is a valid approximation when the typical size of the system is much smaller than the wavelength. The results might be generalized to work for other resonators by using the distributed element model~\cite{pozar_microwave_2011, nazarov_quantum_2009}. However, this can become rather complicated depending on if it is necessary to calculate the interaction between resonator and spin using the retarded time (Jefimenko's equations). Moreover, it could be that the interaction depends on the current density rather than the current, all of which is outside the scope of this paper. We conveniently  describe a series RLC circuit, see Fig.~\ref{fig:figure1}b.\\
As there is a direct analogy with the mechanical resonator, it is straightforward to write down the complete Lagrangian and derive the EOM. From this we calculate something similar to the susceptibility, but more commonly used in electromagnetism, the impedance $Z(\omega)\equiv\frac{V(\omega)}{\tilde{I}(\omega)}$.\\
The electromagnetic analog of the displacement $q$ is the charge $Q_e$. However, instead of writing down $Q_e$ and `momentum variable' $\dot{Q}_e$, we prefer to work with the current $I\equiv\dot{Q}_e$. The `position variable' $Q_e$ then becomes $\int\!\!\mathrm{d}tI$. This results in the RLC-resonator's EOM as
\begin{equation}
L\dot{I}+RI+\frac1C\int\!\! dtI+\frac{\mathrm{d}}{\mathrm{d}t}\left(\bm{\mu}\cdot\frac{\partial}{\partial I}\bm{B}\right)=V(t).
\end{equation}
The resulting interaction term is slightly different compared to that of the mechanical resonator. The zeroth order term vanishes conveniently due to the time derivative, leading to the impedance interaction term $z(\omega)=-i\omega\frac{\mathcal{F}\left\{-\bm{\mu}\cdot\frac{\partial}{\partial I}\bm{B}\right\}}{\tilde{I}(\omega)}$. The spin's EOM does not change, apart from change of variable $q\to I$. This results in an extra impedance $z=z_1+z_2+z_3$, equivalent to the $\kappa$-terms, where
\begin{align}
z_1&=i\omega\mu_s\tanh\left(\beta\mu_sB_0\right)\left|\bm{B}''_{\parallel\bm{\hat{B}}_0}\right|,\\
z_2&=i\omega\frac{\mu_s}{B_0}\frac{\beta\mu_sB_0}{\cosh^2\left(\beta\mu_sB_0\right)}\left|\bm{\hat{B}}'_{\parallel\bm{\hat{B}}_0}\right|^2\frac{1}{1+i\omega T_1},\\
z_3&=i\omega\frac{\mu_s}{B_0}\tanh\left(\beta\mu_sB_0\right)\left|\bm{B}'_{\perp\bm{\hat{B}}_0}\right|^2\cdot\nonumber\\
&\qquad\left(1-\frac{2\frac{T_2}{T_1}-\left(\omega T_2\right)^2+i\omega T_2\left(1+2\frac{T_2}{T_1}\right)}{\left(1+i\omega T_2\right)^2+\left(\omega_sT_2\right)^2}\right).
\end{align}
The resonators complex impedance then becomes
\begin{equation}
Z(\omega)=i\omega L + R + \frac{1}{i\omega C} + z.
\end{equation}
It is much harder to simplify the $z$-terms as done in Sect.~\ref{sec:bath} when partially integrating over a whole sample because $I$ is, unlike $q$, not a Cartesian direction. However, one thing simplifies the $z$ term reasonably: the law of Biot-Savart shows a linear dependence on $I$ implying that $z_1$ vanishes. Note that it is very well possible that the frequencies of interest are comparable to $\frac{1}{T_2}$ or $\omega_s$. In this case one should calculate the whole term. Moreover one should be careful with the implied condition of Eq.~\ref{eq:condition}, i.e. $\pi\gamma_s^2\left|\bm{B}'\right|^2I^2T_1g\left(\omega\right)\ll1$ when probing the resonator.\\
In Fig.~\ref{fig:figure4} we provide an example of an electromagnetic RLC-circuit fabricated on top of a silicon sample with a native oxide. The electron spins inside the native oxide couple to the inductor changing the resonators resonance frequency and Q-factor.

\section{Resonator coupling to other systems}
\label{sec:gen}
So far we have done nothing more than rigorous math to calculate the susceptibility of a system were the physical process is precisely known. However, the physical nature of the interaction between a resonator and a general two level systems (2LSs) can be different from the simple magnetic field interaction and will often even be unknown. This subject has been studied in glassy systems long before it found its application in quantum technology~\cite{phillips_tunneling_1972}. The field revived when it was found in experiments that the electric permittivity and loss factor of a nonmagnetic glass do actually depend on the magnetic field~\cite{strehlow_magnetic_2000}. It was only until recently, around the same time as this paper appeared on a preprint server, that \citet{jug_realistic_2016} provided an intuitive and elegant explanation based on a $\bm{B}$-field dependent density of states and heat capacity. Indeed, expanding the average energy term, as we did in App.~\ref{sec:appB}, leads to the heat capacity which resulted in $\kappa_2\propto z_2\propto \frac{x}{\cosh^2(x)}$ with $x$ the Zeeman energy $-$ temperature ratio. These similar results in combination with the results obtained in this paper imply two things: 
First the $\bm{B}$ does not have to be the physical magnetic field. It is always possible to rewrite the two state Hamiltonian to
\begin{align}
H=E_0+\frac{\epsilon}{2B_0}\bm{\sigma}\cdot\left(\bm{B}_0-q\bm{B}'+q^2\bm{B}''+\ldots\right),
\end{align}
where $\bm{B}$ can be any field that splits the energy levels, leading to an energy difference $\epsilon$ when $q=0$. Here $E_0$ is an uninteresting energy-offset and $\bm{\sigma}$ is a vector containing the Pauli matrices. The interaction strength is determined by $\frac{\partial}{\partial q}\bm{B}$, hence it is important that $\bm{B}$ depend on $q$, which is the generalized coordinate of the mechanical resonator, or generalized velocity of the electromagnetic resonator. Because the expectation values of the Pauli matrices $\bm{\sigma}$ are described by the Bloch equations, the derivations in this paper apply to any resonator-2LS system. Just substitute $\mu_s\to\frac{\epsilon}{2B_0}$ into the $\kappa$ and $z$ terms.\\
Secondly, this result can be easily generalized to a system with $2S+1$ energy levels (with $S$ an integer or half integer) by expanding the Brillouin function from App.~\ref{sec:appB} and substituting
\begin{align}
\tanh\left(\beta\mu_sB_0\right)&\to(2S+1)\coth\left((S+\textstyle{\frac12})\beta\epsilon\right)-\coth\left(\textstyle{\frac12}\beta\epsilon\right)\\
\frac{\beta\mu_sB_0}{\cosh^2\left(\beta\mu_sB_0\right)}&\to\frac{-\textstyle{\frac12}(2S+1)^2\beta\epsilon}{\sinh^2\left((S+\textstyle{\frac12})\beta\epsilon\right)}+\frac{\textstyle{\frac12}\beta\epsilon}{\sinh^2\left(\textstyle{\frac12}\beta\epsilon\right)}
\end{align}
into the $\kappa$ and $z$-terms. This $2S+1$-state quantum system must be isomorphic to a spin-$S$ particle and hence meet two conditions: 1) the energy levels are equally spaced and 2) transitions are only possible to adjacent energy levels.\\

\section{Discussion and Conclusions}
\label{sec:discon}
We have calculated the linear response function of a mechanical and electromagnetic resonator coupled to a spin. The linear response function of the resonator shows extra terms that result in a shift of the resonance frequency and a drop of the Q-factor of the resonator, compared to the bare resonator characteristics. Moreover, we have generalized these results to the coupling with an energy level system with an arbitrary amount of equally spaced energy levels. 
In practice this means that despite having nonmagnetic samples and frequencies that are not even close to the Larmor frequency, one encounters dissipation of the resonator due to the inhomogeneous field it creates. Eventually this might not be a surprise since the resonator alters the heat capacity of the spin's equivalent spin ensemble. Although this is closely related to the magnetic loss enhancement in nonmagnetic glassy systems~\cite{jug_realistic_2016}, we did not find any description in literature that provides a quantitative and detailed account of how this influences the linear response of the resonator, despite the many reported and unexplained results~\cite{imboden_evidence_2009,venkatesan_dissipation_2010,bruno_reducing_2015}. The results presented here have been experimentally verified~\cite{den_haan_spin-mediated_2015} and have been used to calculate the frequency shift in a simple, yet powerful, saturation measurement protocol~\cite{wagenaar_probing_2016}.\\
We have chosen to do the calculations completely in the (semi)classical regime as we are especially interested in the expectation value of spin and resonator. Moreover this leads to an intuitive description and fairly simple calculations. The classical treatment has it limitations though: \citet{berman_magnetic_2006} have raised the point that, if the cantilever position is constantly measured, there is an influence on the spin because of the projections that are constantly occurring in the act of measuring. This might introduce random quantum jumps which, when they are not time averaged over timescales longer than $T_1$, are not taken into account in our description. Furthermore, when pulses are applied, for example in spin resonance techniques, a precise time evolution of the system is needed. Moreover, sending hard pulses might violate the condition for the temperature and linear response of the spin that we have encountered in Sec.~\ref{sec:bas}. In this case one might move to a calculation involving the spin-operators. The theory presented here would still give a fair indication about the enhancement of dissipation, which is of importance in the field of hybrid quantum systems that are pushing the limit of macroscopic superpositions\cite{lee_topical_2016,wezel_nanoscale_2012}. 

\begin{acknowledgments}
We thank M. de Wit and G. Welker for discussions and proofreading this manuscript. This work is part of the single phonon nanomechanics project of the Foundation for Fundamental Research on Matter (FOM), which is part of the Netherlands Organisation for Scientific Research (NWO). 
\end{acknowledgments}

\bibliography{References}
\widetext
\appendix

\section{Resonator - Semiclassical Spin Lagrangian}
\label{sec:appA}
The semiclassical magnetic moment $\bm{\mu}$ can be seen as a vector with an azimuth $\phi$ and a polar angle $\theta$, where the poles of the spherical coordinate system ($\theta=0^\circ$ and $180^\circ$) lie on the axis parallel to the magnetic field. $\theta$ and $\dot{\phi}$ can be seen as the two degrees of freedom that a spin has. Then the Lagrangian $L=\bm{\mu}\cdot\bm{B}(q) + S\hbar\dot\phi\cos\theta$ reveals the Bloch equations for a spin-$S$ particle, but then without decay and for magnetic moment instead of magnetization. The last term of the Lagrangian describes the internal dynamics of the spin. Substituting this into the full Lagrangian in Eq.~\ref{eq:LagS}, we find
\begin{align}
L = \frac12 m\dot{q}^2-\frac12kq^2+\bm{\mu}\cdot\bm{B}(q) + S\hbar\dot\phi\cos\theta.
\end{align}

\section{Equilibrium magnetic moment}
\label{sec:appB}
By definition of the equilibrium vector we can state that $-\bm{\mu}_\infty\cdot\bm{B}=\langle E\rangle$, where $\langle E\rangle$ is the equivalent ensemble average for the energy, or for a single spin the averaged energy over all the points in time with equal $q$. The limited energy levels make it easy to calculate the average energy: For spin-S there are $2S+1$ energy levels with energies $E_k = -kg_s\mu_s\left|\bm{B}\right|$ with $k = -S,-S+1,\ldots,S$. Using the relation between internal energy and the canonical partition function, this results in
\begin{align}
\label{eg:muinf_original}
\bm{\mu}_\infty &= \mu_s\Big(\left(2S+1\right)\coth\big(\left(2S+1\right)\beta\mu_s\left|\bm{B}\right|\big)-\coth\big(\beta\mu_s\left|\bm{B}\right|\big)\Big)\bm{\hat{B}}\\
&\!\!\!\stackrel{S=\frac12}{=} \mu_s\tanh\left(\beta\mu_s\left|\bm{B}\right|\right)\bm{\hat{B}}.
\end{align}
This result is also known as the Brillouin function for the Zeeman energy. The imposed direction $\bm{\hat{B}}$ follows from Curie's law.  The result might be different when the spin has (strong) interaction with its neighbors and when this leads to anisotropic effects, although some of these effects might be included in the $q$ independent part of $\bm{B}$.

\section{Zeroth order solution}
\label{sec:appC}
If the magnetic field generated by the oscillating magnet is given by $\bm{B}(\bm{r})$ in the magnet's rest frame, then in the laboratory frame the magnetic field is $\bm{B}(\bm{r}-\lambda\bm{q})$. Around the spin position $\bm{r}_s$ the magnetic field is 
\begin{align}
\label{eq:Bexpansion}
\bm{B}=\bm{B}_0-q\bm{B}'+\frac12q^2\bm{B}''+\ldots.
\end{align}
Here $\bm{B}_0\equiv\bm{B}(\bm{r}_s)$, $\bm{B}'\equiv\left.\frac{\partial\bm{B}}{\partial q}\right|_{\bm{r}=\bm{r}_s}$ and $\bm{B}''\equiv\left.\frac{\partial^2\bm{B}}{\partial^2 q}\right|_{\bm{r}=\bm{r}_s}$.\\
Next we substitute $q\to\lambda q$ and expand $\bm{\mu}_\infty$ for spin-$\frac12$ up to first order in $\lambda$ and omit higher order terms
\begin{align}
\bm{\mu}_\infty &= \mu_s\tanh\left(\beta\mu_sB_0\right)\bm{\hat{B}}_0 - q\left(\tanh\left(\beta\mu_sB_0\right)P_\perp + \frac{\beta\mu_sB_0}{\cosh^2\left(\beta\mu_sB_0\right)}P_\parallel\right)\frac{\bm{B}'}{B_0},
\end{align}
where $P_\parallel$ and $P_\perp$ are projections parallel and perpendicular to the $\bm{B}_0$ field respectively, i.e. $P_\parallel \equiv \bm{\hat{B}}_0\bm{\hat{B}}_0^T$ and $P_\perp\equiv\mathbb{1}-\bm{\hat{B}}_0\bm{\hat{B}}_0^T$.
We also set $q\to\lambda q$ into Eqs.~\ref{eq:Bloch} and \ref{eq:Bexpansion} and set $\lambda\to0$ to get the differential equation to solve for $\bm{\mu}_0$:
\begin{align}
\label{eq:diffmu0}
\dot{\bm{\mu}}_0=\left(\gamma_sB_{0\times}-\frac{1}{T_2}P_\perp-\frac{1}{T_2}P_\parallel\right)\bm{\mu}_0+\frac{\mu_s}{T_1}\tanh\left(\beta\mu_sB_0\right)\bm{\hat{B}}_0,
\end{align}
where the $\times$ subscript denotes an antisymmetric matrix such that $A_\times\bm{v}\equiv\bm{v}\times\bm{A}$ for any vector $\bm{v}$ and $\bm{A}$.\\\\
Let $\bm{M}(s)\equiv\int_0^\infty \mathrm{e}^{-st}\bm{\mu}(t)\,\mathrm{d}t$ be the Laplace transform of the magnetic moment and apply the necessary linear algebra to get
\begin{align}
\label{eq:solmu0laplace}
 \bm{M}_0(s) &= \left(\frac{\left(s+\frac{1}{T_2}\right)P_\perp+\omega_s\hat{B}_{0\times}}{\left(s+\frac{1}{T_2}\right)^2+\omega_s^2} + \frac{P_\parallel}{s+\frac{1}{T_1}}\right)\left(\frac1s\frac{\mu_s}{T_1}\tanh\left(\beta\mu_sB_0\right)\bm{\hat{B}}_0+\bm{\mu}(0)\right),
\end{align}
with $\omega_s\equiv\gamma_sB_0$. The inverse Laplace transform yields the general solution for $\bm{\mu}_0$ in the time-domain:
\begin{align}
\bm{\mu}_0(t) = \mu_s\left(1-\mathrm{e}^{-t/T_1}\right)\tanh\left(\beta\mu_sB_0\right)\bm{\hat{B}}_0+\begin{pmatrix}
  \mathrm{e}^{-t/T_2}\cos(\omega_st) & -\mathrm{e}^{-t/T_2}\sin(\omega_st) & 0\\
  \mathrm{e}^{-t/T_2}\sin(\omega_st) & \mathrm{e}^{-t/T_2}\cos(\omega_st) & 0 \\
  0 & 0 & \mathrm{e}^{-t/T_1} 
 \end{pmatrix}\bm{\mu}(0).
\end{align}\\
To retrieve some intuition for the results we choose to present the last term as a matrix which is given in a non-rotating Cartesian basis with $\bm{\hat{z}}=\bm{\hat{B}}_0$.\\
To estimate the static displacement we use $\bm{\mu}_0(\infty)$, which is of course the same as $\bm{\mu}_\infty(q=0)$, to find the change of equilibrium position
\begin{align}
q\to q-\frac{\bm{\mu}_0\cdot\bm{B}'}{k+\delta k}\approx q - \frac{\mu_s}{k}\tanh\left(\beta\mu_sB_0\right)\bm{\hat{B}}_0\cdot\bm{B}',
\end{align}
where in the last step we neglected $\delta k$, the effective extra stiffness coming from the terms linear in $q$. 

\section{First order solution}
\label{sec:appD}
As argued in the main text, we can ignore the terms that decay or depend on initial conditions. As a consequence we can take $\bm{\mu}_0=\mu_s\tanh\left(\beta\mu_sB_0\right)\bm{\hat{B}}_0$. This leads immediately to one of the interaction terms. Taking $\mathcal{F}\left\{-q\bm{\mu}_0\cdot\bm{B}''\right\}=\kappa_1\tilde{q}(\omega)$ with $\tilde{q}(\omega)=\mathcal{F}\{q(t)\}$ we arrive at 
\begin{align}
\kappa_1 = -\mu_s\left|\bm{B}''_{\parallel\bm{\hat{B}}_0}\right|\tanh\left(\beta\mu_sB_0\right),
\end{align}
where $\left|\bm{B}''_{\parallel\bm{\hat{B}}_0}\right|=\bm{B}''\cdot\bm{\hat{B}}_0$.\\
Next, we need to find $\bm{\mu}_1$. Again this is done by substituting $q\to\lambda q$ and extracting the terms that are linear in $\lambda$ only. We find
\begin{align}
\dot{\bm{\mu}}_1	&=\left(\gamma_sB_{0\times}-\frac{1}{T_2}P_\perp-\frac{1}{T_2}P_\parallel\right)\bm{\mu}_1\nonumber\\
				&\qquad+q(t)\left(\left(\frac{1}{T_2}-\frac{1}{T_1}\right)C-\gamma_sB'_\times\right)\bm{\mu}_0\nonumber\\
				&\qquad-q(t)\frac{\mu_s}{B_0T_1}\left(\tanh\left(\beta\mu_sB_0\right)P_\perp+\frac{\beta\mu_sB_0}{\cosh^2\left(\beta\mu_sB_0\right)}\right)\bm{\hat{B}}',
\end{align}
where $C\equiv\frac{1}{B_0}\left(\bm{\hat{B}}_0\bm{B}'^TP_\perp+P_\perp\bm{B}'\bm{\hat{B}}_0^T\right)$.\\
The first line is the same as in Eq.~\ref{eq:diffmu0} and therefore leads to the same matrix as in Eq.~\ref{eq:solmu0laplace} using the same non-rotating Cartesian basis with $\bm{z}=\bm{\hat{B}}_0$. This leads to
\begin{align}
\bm{M}_1(s) &= \left(\frac{\left(s+\frac{1}{T_2}\right)P_\perp+\omega_s\hat{B}_{0\times}}{\left(s+\frac{1}{T_2}\right)^2+\omega_s^2} + \frac{P_\parallel}{s+\frac{1}{T_1}}\right)\cdot\nonumber\\
 &\qquad\left( \tanh\left(\beta\mu_sB_0\right) \left( \left(\frac{1}{T_2}-\frac{2}{T_1}\right) P_\perp\bm{B}'-\omega_s\bm{\hat{B}}_0\times\bm{B}' \right) - \frac{\beta\mu_sB_0}{\cosh^2\left(\beta\mu_sB_0\right)}P_\parallel\bm{B}'\right)\frac{\mu_s}{B_0T_1}Q(s),
\end{align}
with $\bm{M}_1(s)$ and $Q(s)$ being the Laplace transform of $\bm{\mu}_1(t)$ and $q(t)$ respectively.\\
$\bm{M}_1$, and thus $\bm{\mu}_1$, can be easily split in a part that is parallel and perpendicular to $\bm{\hat{B}}_0$. It follows from Eq.~\ref{eq:inttermexpansion} that we need specifically the product $\bm{\mu}_1\cdot\bm{B}'$ for the interaction term. So let us write $\mathcal{F}\left\{\bm{\mu}_1\cdot\bm{B}'\right\} = \tilde{q}(\omega)\left(\kappa_2+\kappa_3\right)$ where $\kappa_2$ and $\kappa_3$ come from the parallel and perpendicular parts of $\bm{\mu}_1$ respectively. Finally we move to the Fourier domain, which is possible since all poles lie in the Re$(s)<0$ regime. This leads to
\begin{align}
\kappa_2 = -\frac{\mu_s}{B_0}\left|\bm{\hat{B}}'_{\parallel\bm{\hat{B}}_0}\right|^2\frac{\beta\mu_sB_0}{\cosh^2\left(\beta\mu_sB_0\right)}\frac{1}{1+i\omega T_1},
\end{align}
where $\left|\bm{\hat{B}}'_{\parallel\bm{\hat{B}}_0}\right|^2=\bm{B}'^TP_\parallel\bm{B}'$.\\
For $\kappa_3$ we find
\begin{align}
\kappa_3 = -\frac{\mu_s}{B_0}\left|\bm{B}'_{\perp\bm{\hat{B}}_0}\right|^2\tanh\left(\beta\mu_sB_0\right)\left(1-\frac{2\frac{T_2}{T_1}-\left(\omega T_2\right)^2+i\omega T_2\left(1+2\frac{T_2}{T_1}\right)}{\left(1+i\omega T_2\right)^2+\left(\omega_sT_2\right)^2}\right),
\end{align}
where $\left|\bm{B}'_{\perp\bm{\hat{B}}_0}\right|^2=\bm{B}'^TP_\perp\bm{B}'$.

\end{document}